# Cell membrane disruption by vertical nanopillars: the role of membrane bending and traction forces.

*Rosario Capozza[a], Valeria Caprettini[a,b], Carlo A. Gonano[a], Alessandro Bosca[a], Fabio Moia[a], Francesca Santoro[c], Francesco De Angelis[a]\**

[a]Istituto Italiano di Tecnologia, via Morego 30, 16163 Genova, Italy

[b]Università degli studi di Genova, Genova, Italy

[c]Center for Advanced Biomaterials for Healthcare, Istituto Italiano di Tecnologia, 80125 Napoli, Italy

\*corresponding author's email: francesco.deangelis@iit.it

KEYWORDS: nanopillars, lipid bilayer, nanopore, membrane disruption, molecular dynamics, intracellular delivery

## ABSTRACT

Gaining access to the cell interior is fundamental for many applications, such as electrical recording, drug and biomolecular delivery. A very promising technique consists of culturing cells on nano/micro pillars. The tight adhesion and high local deformation of cells in contact with nanostructures can promote the permeabilization of lipids at the plasma membrane, providing access to the internal compartment. However, there is still much experimental controversy regarding when and how the intracellular environment is targeted and the role of the geometry and interactions with surfaces. Consequently, we investigated, by coarse-grained molecular dynamics simulations of the cell membrane, the mechanical properties of the lipid bilayer under high strain and bending conditions. We found out that a high curvature of the lipid bilayer dramatically lowers the traction force necessary to achieve membrane rupture. Afterwards, we experimentally studied the permeabilization rate of cell membrane by pillars with comparable aspect ratios but different sharpness values at the edges. The experimental data support the simulation results: even pillars with diameters in the micron range may cause local membrane disruption when their edges are sufficiently sharp. Therefore, the permeabilization likelihood is connected to the local geometric features of the pillars rather than diameter or aspect ratio. The present study can also provide significant contributions to the design of 3D biointerfaces for tissue engineering and cellular growth.

**INTRODUCTION**

Direct access to the intracellular compartment is an open challenge [1] with many potential applications, such as gene transfection [2, 3], biomolecule delivery [4], and electrical recording in electroactive cells [5]. The main difficulties are related to the impermeability of the plasma membrane, which, after a billion years of evolutionary defenses, strictly controls the trafficking in and out of the cell. The most popular methods for intracellular delivery are electroporation [6], chemical transfection, and virus-mediated transduction, although novel methods have been considered recently [7, 8, 9, 10]. Among them, local membrane permeabilization through vertical nanopillars or other 3D nanostructures is emerging as a robust approach [11, 12, 13, 14, 15, 16, 17]. Essentially, the concept relies on arrays of vertical standing nanostructures in a fakir bed-like configuration. When cells are cultured on these arrays, the plasma membrane exhibits tight adhesion to the pillars or even engulfment-like events [18]. These processes often lead to a spontaneous increase in membrane permeability [19] that can be used to deliver molecules into the cytosol by bypassing the conventional biochemical pathways or to record intracellular electrical activity via the enhanced electrical coupling between the conductive pillar and the cell [5]. In general, there is a wide interest in designing novel 3D bio-interfaces for tissue engineering and cellular growth [20] to investigate how cells interact and proliferate onto these types of geometries [21, 22] and which structures improve the cell viability [23]. However, the exact mechanism enabling local permeabilization is still not fully understood, and many controversies still exist. Recent studies suggested that the internalization of molecules is not due to the temporary disruption of the plasma membrane in contact with the 3D nanostructures but is instead driven by membrane deformation [24] or enhancement of the clathrin-mediated endocytosis process of the cell at the interface with sharp edges [25]. On the contrary, a previous study, developing a simple but effective mechanical continuum model of elastic cell membranes [26], ascribes such complex behaviors to actual penetration of the plasma membrane. In that work, two main mechanisms at the interface are considered, namely "impaling", where cells land on a bed of nanowires, and "adhesion-mediated" permeabilization, which occurs as cells spread on the substrate and generate adhesion force. In the former, the force leading to the membrane disruption is gravity, whereas in the latter mechanism, this force is the adhesion-force provided by membrane proteins. In both cases, membrane permeabilization occurs when the nanopillar generates sufficient tension to overcome a critical membrane tension value. However, the pillar is modeled as a cylindrical probe with a hemispherical tip, and the effect of local geometry (e.g., the sharpness of pillar edge) is not taken into account. Hence, only the diameter, the height and the spacing between the pillars determine the penetration forces for a cell line of a given stiffness. This approach may explain the experimental reports of spontaneous poration observed in vertical nanostructures of small diameters on the order of 50-200 nm [27, 28, 29] or below. Indeed, as we will subsequently show, we found experimentally that spontaneous permeabilization may occur even for much larger pillar diameters of approximately 2 μm. Such a finding is also confirmed by another study that recently reported spontaneous membrane disruption by pillars of 1 μm in diameter, even in absence of adhesion with the substrate [30]; hence, the models mentioned above cannot fully explain the membrane permeabilization. In other words, the reasons why these large nanopillars can effectively permeabilize the membrane are still unclear, thus demanding a more complex scenario that includes not just gravity and adhesion but also traction forces, membrane deformations (bending) and local geometric features (sharp edges on the pillars). Obviously, biological mechanisms and surface properties also participate in the increase in cell internalization processes, but the present mechanical model will not take them into account.

Molecular dynamics simulation is a powerful tool that may elucidate the behavior of these systems. However, to the best of our knowledge, there are still few molecular dynamics studies investigating the mechanical properties of a membrane when in contact with a nanostructure [31]. Furthermore, most molecular dynamics studies tackle the problem only in the regime of small bending deformations [32, 33, 34], which is not applicable for this case of study.

In this work, we first undertake molecular dynamics simulations in synergic combination with a mechanical model of the cell membrane that goes beyond the linear response approximation. We found that the bending of the membrane is characterized by an elastic regime at low bending angles, followed by a plastic one at higher angles.

We also included the local geometry of the pillar by investigating the role of curvature or edges at the pillar tip (different from the pillar diameter), and we found that a high curvature favors the rupture even at very low tensile strength.

Afterwards, we show that simulation results are supported by experimental data suggesting that when cells are cultured on pillars with diameters in the micrometric range, their permeabilization likelihood is strongly increased in the presence of sharp edges. We conclude that, under the given conditions, the local curvature may dramatically affect the lipid bilayer permeabilization to a greater extent than the effect of the pillar diameter.

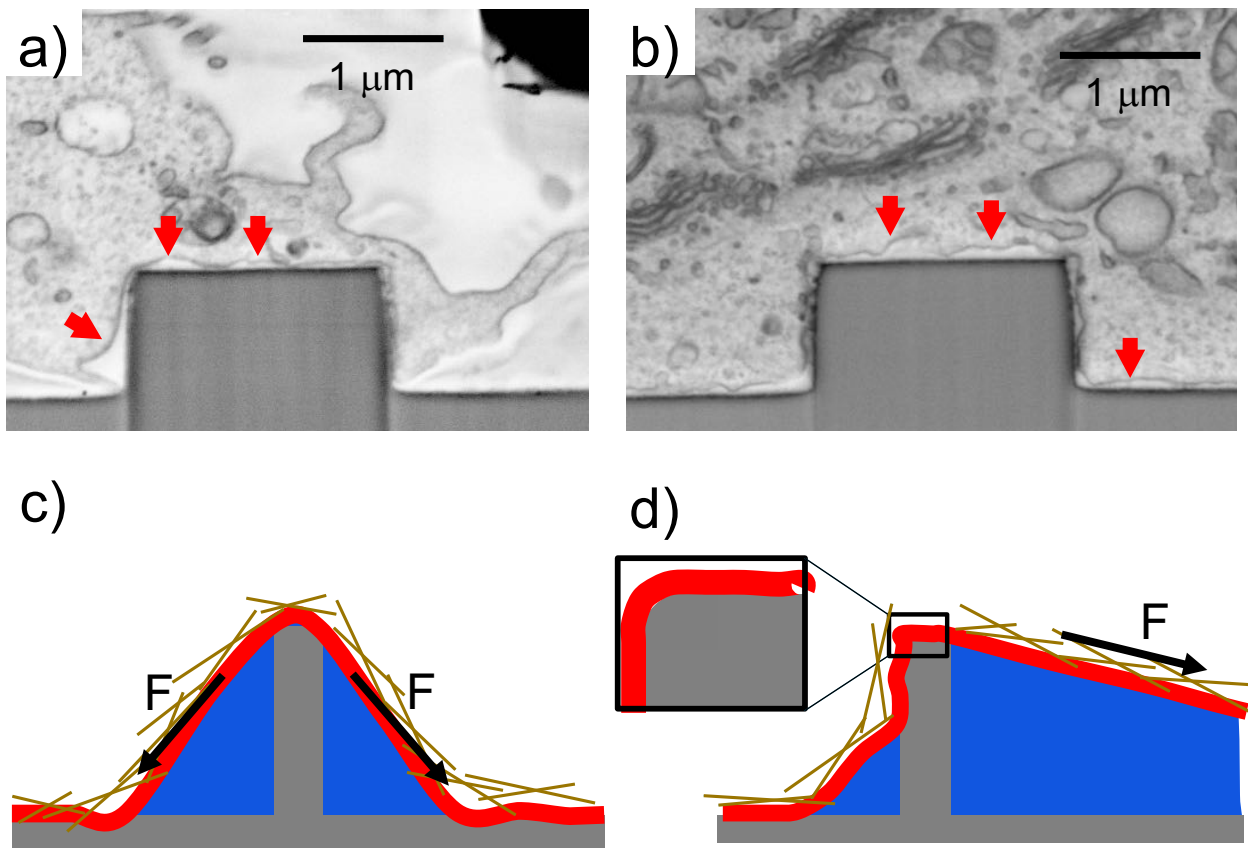

**Figure 1. SEM images showing a FIB cross sections of a cell cultured on an array of 2μm diameter nanopillars. The samples were fixed and stained following a recently developed ROTO protocol (see S2 for more details) and embedded in a thin film of epoxy resin. The cell membrane (indicated by red arrows) can assume a "tent-like" configuration (a) or be tightly wrapped around the pillars (b). c) Adhesion-induced**

**mechanism of permeabilization. d) Traction-induced mechanism of permeabilization. Inset: the membrane's bending is dictated by the local radius of curvature rather than by the nanopillar diameter.**

**RESULTS AND DISCUSSION**

Figure 1 shows a SEM/FIB cross-sectional image of cell cultured on nanopillars with a diameter of 2 μm and height of 1.5 μm, obtained by using a technique explained elsewhere [35]. This image displays possible configurations of the cell/nanostructure interface, and highlights the proximity of the cell membrane to the pillars: the cell can tightly wrap around the 3D pillar (b) or assume a "tent-like" configuration (a). In both cases, the cell membrane conforms to the pillar's head geometry, where we assume most of tensile strength is concentrated, via adhesion with the substrate and forces exerted through the cell body. Therefore, the membrane curvature is dictated by the pillar's edge. Furthermore, the forces acting on the membrane can be oriented in any direction depending on the local configuration of the system, as sketched in panels c) and d) of Figure 1. Traction and bending cannot be considered separately, and they may cooperate to lower the threshold for local permeabilization or nanopore creation. For instance, when the traction force is mainly directed laterally (parallel to the substrate), the pillar may behave like a knife edge, meaning that only the size of the edge matters and the diameter plays a minor role.

**Simulation model.** We used a 2D coarse-grained model of the cell membrane in which the lipid molecule is made of a hydrophilic Lennard-Jones (L-J) particle (the head) and a hydrophobic tail made of five L-J beads (Figure 2b). We included an additional harmonic interaction between the L-J beads of the tail. The water is simulated as a single L-J bead [36]. First, we tested the validity of our system by simulating free lipid molecules randomly dispersed in water and checking that the lipids undergo a self-assembly process that leads to the formation of the lipid membrane. As expected, under the conditions we used, a bilayer is formed spontaneously with the hydrophilic heads pointing toward water and the tails clustering together, minimizing their interaction with water (see S3). The average distance between hydrophilic heads is on the order of experimental values [37] (~0.8 nm). Clearly, with a 2D model, we cannot reproduce all of the 3D phenomena, for example, the lipid diffusion through the membrane's surface.

In addition, the molecular dynamics model presented here considers just the lipid bilayer and not the cytoskeleton and membrane proteins that certainly affect the mechanical properties of the cell. However, we carefully checked that our numerical simulations can qualitatively reproduce many real processes (self-assembly, bilayer, micelles, vesicles, etc.) that depend on the geometrical features of molecules [37]. This method can be regarded as a good compromise between the computational cost and the accuracy of the mechanical analysis and allows a statistical analysis of results. After the bilayer formation, we replicate the structure to create a system size at will and set up the initial configuration (Figure 2a). The typical size of our system is on the order of 100 nm.

**Modeling of the plasma membrane mechanical response.** The mechanical properties of a material are often described by means of a mathematical relation linking the strain $\varepsilon$ (ratio between a displacement and a rest length) to the applied stress $\sigma$ (force per unit area). For an ideal linear material, the applied stress is given by $\sigma = E \cdot \varepsilon$, where $E$ is the Young's modulus (see S4). The Young's modulus $E$ is a characteristic of the material, so it is an *intensive* parameter in the sense that it does not depend on extensive properties of the considered object (e.g., length and mass); however, membrane deformation is often described in terms of stiffness and applied forces [26, 33, 37]. For example, the area stretch modulus $K_A$ (characterizing the rigidity of a membrane to traction force) and the bending stiffness $K_{bend}$ (linking the curvature to the applied moment) are extensive parameters because they depend on the membrane's length and thickness (see S5). We decided to calculate both intensive and extensive quantities. Since the cell membrane may undergo large deformation when in contact with nanostructures, the approximation of a linear material is no longer valid, and the membrane could also exhibit plastic behavior (hysteresis). In this case, the Young's modulus can be interpreted as the derivative of the stress $\sigma$ with respect to the strain $\varepsilon$, formally:

$$\Delta\sigma = E(\sigma, \varepsilon) \cdot \Delta\varepsilon$$

More simply, extracting the strain-stress relation is a standard strategy for characterizing the elastic properties of a material. For that purpose, we simulated a cell membrane loaded by cylinders measuring the displacements and the applied forces, as shown in Figure 2a.

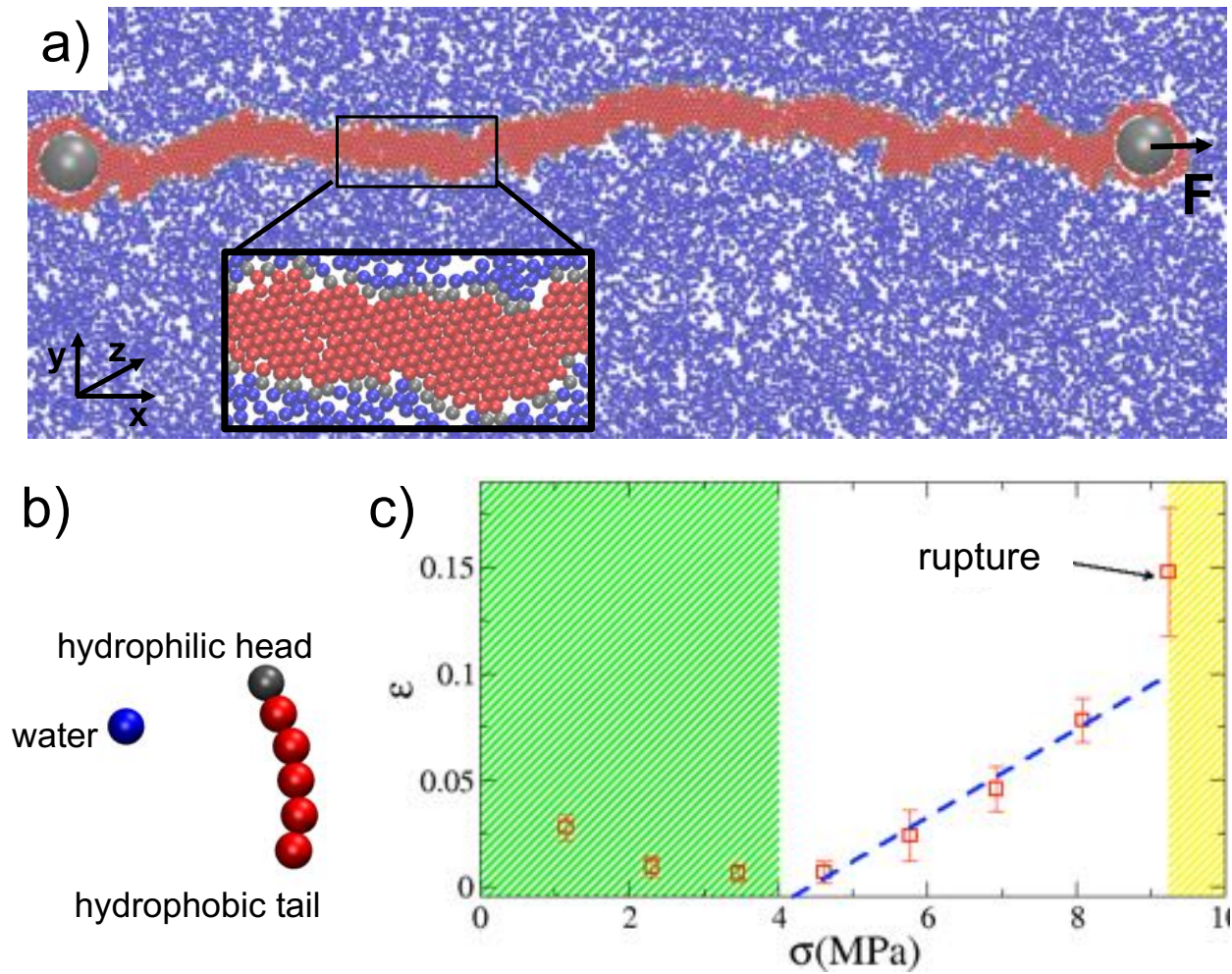

**Figure 2. a)** Snapshot of a simulated segment of membrane subject to a horizontal traction force F. Here, water is indicated in blue, hydrophilic heads in gray and hydrophobic tails in red. Inset: blow-up of the simulated system showing the molecular arrangement in detail. **b)** Coarse-grained model of water and the lipid molecule with a single bead for the hydrophilic head and five beads for the hydrophobic hydrocarbon tail. **c)** Strain-stress curve showing three loading regimes: a thermally dominated regime (thermal noise larger than the applied stress), an elastic regime, and failure for F > 28 pN (corresponding to σ > 9.2 MPa).

Afterwards, we calculated the strain-stress curve through the classical Euler-Bernoulli beam model connecting the bending moment $M_z$ to the deflection of the beam (the membrane) through the Young's modulus $E$ and a local linearization.

In the simple case of pure traction, we calculate the stress-strain curve by applying a traction force to the edges of the membrane (Figure 2a) and estimating the ultimate tensile strength as the force $F_{TS}$ we need to apply to break the membrane. After embedding two cylinders into the membrane segment, we keep one fixed, while the other one applies a force F along the positive direction of the x axis. The force is increased in steps $\Delta F = 3.5\ pN$ and kept constant for a time interval $\Delta t = 1.2\ ns$. During the step-wise increase in force, we monitored the strain $\varepsilon$ of the membrane, and we found that the membrane ruptures at $F = 28\ pN$ (corresponding to an ultimate tensile stress $\sigma = 9.2\ MPa$). Therefore, the value $F_{TS} = 28\ pN$ represents the tensile strength of the membrane in the straight configuration.

The longitudinal strain $\varepsilon_{xx}$ is the ratio between the displacement $\Delta s_x$ and the rest length $l_x = 92$ nm.

$$\varepsilon_{xx} = \frac{\Delta s_x}{l_x}$$

The applied stress $\sigma_{xx}$ here is simply the force on the membrane's section:

$$\sigma_{xx} = \frac{F_x}{A_t} = \frac{F_x}{l_y l_z}$$

The plot of the strain-stress curve (Figure 2c) clearly shows three loading regimes:

- For $0 \leq \sigma \leq 4.5\ MPa$, the stress does not significantly affect the strain. The membrane fluctuates, and the thermal noise overwhelms the effects of the applied force.
- For $4.5\ MPa < \sigma < 9.2\ MPa$, the membrane responds elastically, and we estimate a Young's modulus $E = 50\ MPa$ by linear regression (blue dashed curve in Figure 2c). This value is in the range of experimental measurements [37].
- For $\sigma > 9.2\ MPa$, membrane rupture occurs.

From Young's modulus $E$, we calculate the area-stretch modulus $K_A = \frac{E \cdot A_t}{l_x} \approx 1.5\ mJ/m^2$. As expected, $K_A$ and $F_{TS}$ are approximately two orders of magnitude smaller than the ones calculated for typical free biological cell membranes, which range from $100 - 250\ mJ/m^2$ [26, 37]. In fact, our system is representative of a longitudinal section of a three-dimensional membrane; thus, the domain's depth must be taken into consideration.

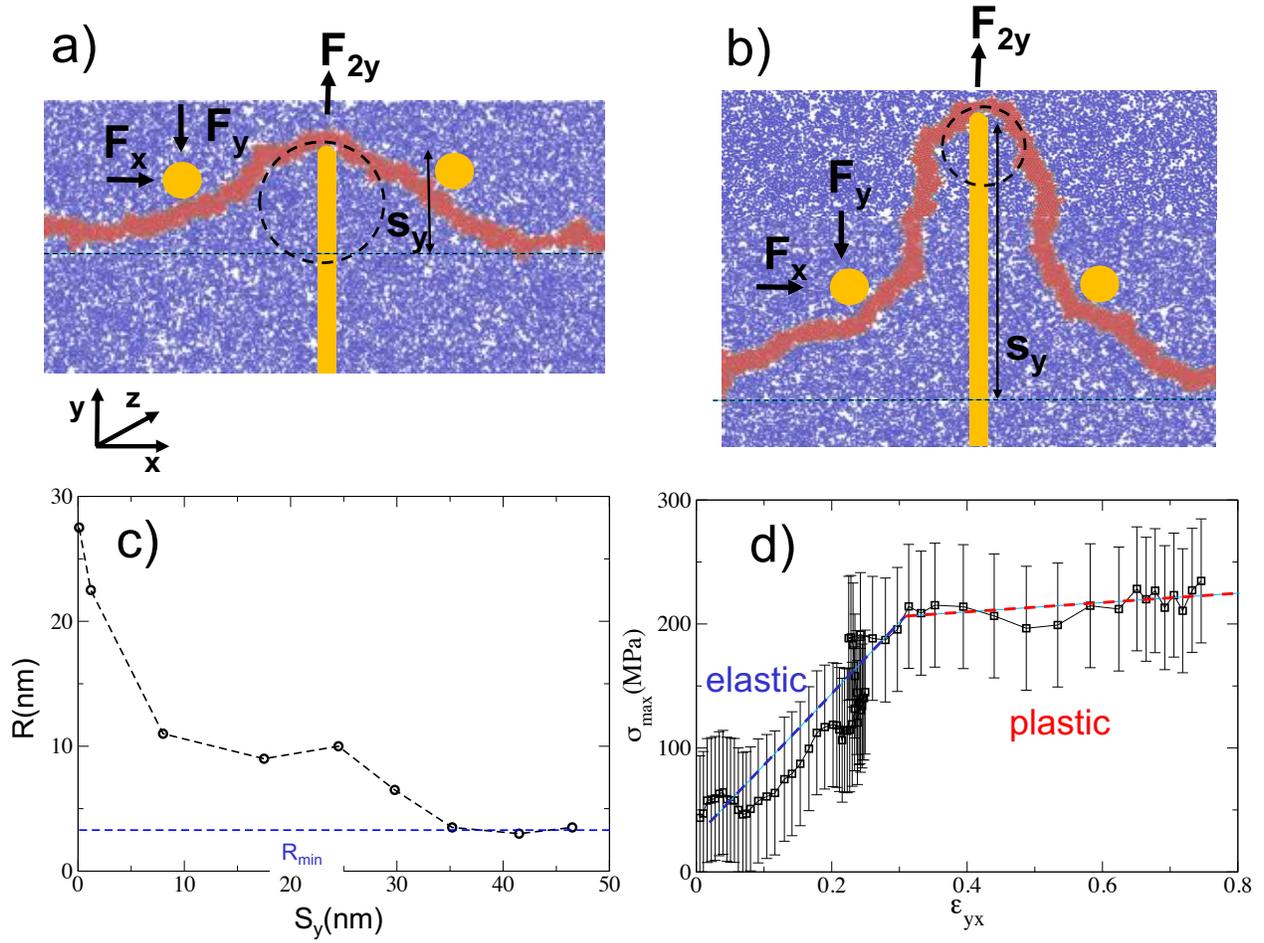

**Figure 3** Configuration of the membrane in response to a small (a) and high (b) deformation $s_y$. c) Behavior of the estimated radius of curvature R as a function of the deformation $s_y$. For $s_y > 35$ nm, the radius of curvature R reaches the minimum $R_{min}$, corresponding to the membrane wrapping around the pillar's tip. d) Stress-strain curve showing the elastic and plastic regimes.

In the following, we describe the calculation of the membrane bending stiffness $K_{bend}$. Previous approaches estimated $K_{bend}$ by analyzing the height (spatial) fluctuation spectrum of a membrane [38, 39, 40, 41]. However, these methods present serious drawbacks due to the slow convergence of long wavelengths [32, 33] and assume a regime of small deformations. In contrast, a membrane in contact with nanopillars could be strongly deformed. For this reason, we studied the bending without invoking neither the hypothesis of small deformations nor the hypothesis of a linear material. Hence, the elastic parameters such as $E$ and $K_{bend}$ could depend on the strain.

We employed a simulation setup similar to the one already used in the literature [32]. A segment of membrane with free ends is pushed in the center through a nanopillar (with a hemispherical tip of radius $r_p \sim 2$ nm) moving upward and two immobile cylinders (Figure 3a, b) interacting just with the hydrophobic tails. The segment is free to flow under the lateral cylinders to avoid stretching and to study the mechanical response under pure bending. During the course of simulations, we measured the $x$ and $y$ components of the force on the cylinders and the pillar and the vertical displacement $s_y$, and we estimate the local radius of curvature R as the best circle approximating the average membrane profile (Figure 3c). For small values of vertical displacement $s_y$, R is independent on the pillar size, and it rapidly decreases reaching the plateau $R_{min} \sim 4$ nm at $s_y \sim 35$ nm (Figure 3c). The value $R_{min}$ represents the smallest possible radius of curvature considering the membrane thickness is $l_y \approx 4.5\ nm$ and corresponds to the membrane wrapping around the pillar's tip. As explained in detail in S6 and S7, for a linear material, the stress $\sigma$ and the local radius of curvature R are both proportional to the bending moment $M_z$. It is then possible to show that the stress-strain relation reads as:

$$\sigma_x = E \cdot \left(\frac{l_y/2}{R}\right) = E \cdot \varepsilon_{xy}$$

where $l_y \approx 4.5$ nm is the average membrane thickness. Since both $\sigma_x$ and the curvature strain $\varepsilon_{xy}$ can be obtained from the simulations, even for a non-linear material, the Young's modulus can be estimated as $E = \frac{\Delta \sigma_x}{\Delta \varepsilon_{xy}}$.

The stress-strain curve is reported in Figure 3d. Two distinct regimes, elastic and plastic, can be clearly identified. In fact, for values of curvature strain $\varepsilon_{xy} < 0.3$, the membrane responds elastically, while for $\varepsilon_{xy} > 0.3$, it is plastically deformed. This latter regime, characterized by a smaller Young's modulus, suggests a hysteretic behavior. In fact, upon retraction of the nanopillar, the membrane does not return to its initial straight configuration, at least in the time interval available to our simulations. Each point in Figure 3d is obtained from a statistical average over ten different simulations. We fitted the points of the stress-strain curve (Figure 3d) with two lines, and we estimated the Young's modulus for the elastic regime $E_{elas} = (600 \pm 150)\ MPa$ and plastic regime $E_{plas} = (75 \pm 25)\ MPa$. Not surprisingly, the value of $E_{plas}$ is (considering the error) equal to the $E$ calculated for the pure stretching, indicating that for $\varepsilon_{xy} > 0.3$, the bending corresponds to an effective stretching of the outer membrane monolayer.

As shown in S5, the relation $K_{bend} = \frac{E}{12} \cdot l_y^3$ allows us to estimate the bending stiffness in the elastic $K_{bend}^{elas} \approx (4.5 \pm 2.6) \cdot 10^{-18} J$ and plastic regime $K_{bend}^{plas} \approx (0.57 \pm 0.37) \cdot 10^{-18} J$. The value of $K_{bend}^{plast}$ is of the same order of magnitude of the ones reported in literature [37], while $K_{bend}^{elast}$ is one order of magnitude larger. This large value of $K_{bend}^{elast}$ can be attributed to the limits of our 2D modeling, which cannot correctly reproduce the interlayer diffusivity of lipid molecules. We stress that the estimation of $K_{bend}^{elast}$ is affected by a large uncertainty related to the dependency on the cube of membrane thickness $l_y$ (see S5) and the one related to Young's modulus $E$.

**Membrane mechanical response to local curvature.** In the following, we investigate how curvature and bending affect the ultimate tensile strength $F_{TS}$. In fact, the edges of 3D nanostructures in contact with the cell membrane could be very sharp, and their sharpness is usually very difficult to control experimentally. The lipid bilayer can adhere to the nanopillar edge and, then, follow its curvature. Traction forces are on the order of 1 nN [18] and in principle too small to promote spontaneous rupture of membrane.

What is the relation between the ultimate tensile strength $F_{TS}$ and the curvature?

To clarify this point, we performed simulations by stretching and bending the membrane simultaneously. The three-cylinder configuration is shown in Figure 4a. Two cylinders (1 and 3) are embedded into the membrane, while cylinder 2 is kept fixed. Cylinder 3 applies a constant force F to the membrane along the positive direction of the x axis and lower than the membrane's ultimate tensile strength, $F_{TS}$ = 28 pN.

Cylinder 1 rotates the membrane around an axis passing through cylinder 2 and directed along z, imposing a curvature on the membrane. We found that the smaller F is, the higher is the curvature (the smaller R) needed to bring the membrane to the rupture, as reported in the snapshots of the system in Figure 4 a, b. In Figure 4c, the local radius R is plotted as a function of $F_{TS}$; each value of R was obtained from a statistical average over ten different simulations. The behavior of the local radius of curvature R as a function of $F_{TS}$ clearly shows that even a tiny force can promote a rupture provided that R is small, i.e., the edge is sharp.

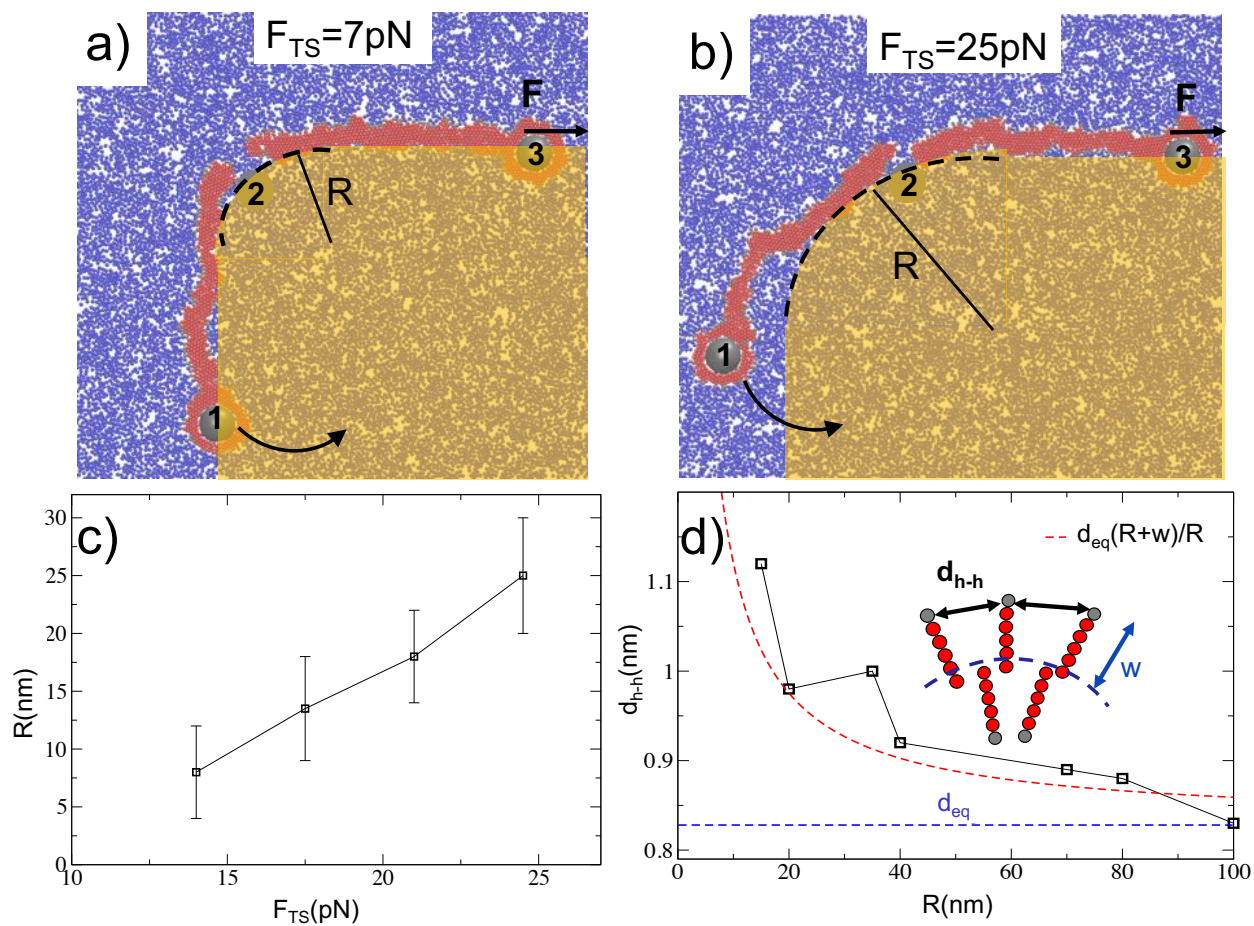

**Figure 4** The three-cylinder simulation geometry used to calculate the effect of the local radius curvature R on the ultimate tensile strength $F_{TS}$. Cylinder 3 applies a constant force F, and cylinder 1 rotates around cylinder 2 (at rest) (a, b). A smaller value of F corresponds to a smaller R needed to bring the membrane to the rupture. The yellow shaded area suggests a possible configuration of the membrane in contact with a nanopillar. c) Behavior of the critical radius of curvature R as a function of $F_{TS}$. d) Average distance between hydrophilic heads, $d_{h-h}$, as a function of the bending radius R, as calculated in the segment where the membrane is bent. The blue dashed line is the equilibrium distance $d_{eq}$ in the absence of bending, and the red dashed line shows the behavior of $d_{h-h}$ from an analytical model. Here, $d_{eq}=0.83$ nm, $w=l_y/2=2.25$ nm.

What is the reason for the curvature-induced membrane weakening?

Figure 4d shows the variation in the average distance between hydrophilic heads, $d_{h-h}$, as a function of the bending radius R, calculated only in the segment where the membrane is bent. In the absence of bending, $d_{h-h}=d_{eq}=0.83$ nm (blue dashed line in Figure 4d), which is the equilibrium distance after the self-assembly process of the membrane (see S3). After the membrane is bent, $d_{h-h}$ deviates from the equilibrium and rapidly increases as the radius of curvature R decreases.

Let us consider an initially straight bar composed of n elements at distance $d_{eq}$ and finite thickness $l_y$ (that is, the average thickness of the membrane). When the bar is bent, the distance $d_{h\text{-}h}$ between the elements of its upper surface (see inset in Figure 4d) increases following the relation $d_{h-h} = \frac{R+(l_y/2)}{R}d_{eq}$ (the red dashed line shown in Figure 4d). The values of $d_{h-h}$ versus $R$, calculated from the numerical simulations, follow approximately the theoretical red dashed line, suggesting that the membrane is effectively deformed like a bar composed of discrete elements.

When $d_{h\text{-}h}$ increases, the membrane is brought out of its equilibrium distance $d_{eq}$, and the probability of nanopore formation increases as well. In this situation, traction forces further increase the probability of defect formation (lowering the activation energy for nanopore formation) and favor breakdown.

**Experimental investigation of permeabilization on sharp pillars.** To confirm the effect of the edge on the cell membrane permeabilization, we fabricated two sets of pillars, denoted as "sharp" and "smooth"; the pillars had an identical pitch of 5 μm, comparable diameter of approximately 2 μm, height of 2.5 μm and 1 μm, respectively, and their most important difference was the edge sharpness, as reported in Figure 5a and 5b (see S1 for more details on the fabrication process). We estimated their radius of curvature as $R_{sharp}\sim 20 \pm 5$ nm for the sharp case and $R_{smooth}\sim 250 \pm 20$ nm for the smooth one from the cross sections of the pillars, as shown in the inset of Figure 2a. Afterwards, we cultured NIH-3T3 cells on these arrays of pillars and administered the impermeable dye propidium iodide in solution together with the permeable calcein AM dye to verify the healthiness of the cells. Surprisingly and contrary to some previous literature results [26], we found that in the case of large but sharp pillars, the dye entered the cell body with a probability of approximately 70%, staining the cells red, as demonstrated by fluorescence images reported in Figure 5c (see SI for more details) maintaining their viability, confirmed by the green stained. These results have been acquired on 400 cells in 6 different cell cultures. In contrast, cells cultured on the smooth pillars showed no sign of dye internalization (no red color) and a green color indicating cell viability but no permeabilization (see Figure 5d). In this last case, the presence of shorter pillars ensures stronger adhesion with the substrate that, according to previous studies [26], is assumed to be one of the factors increasing the permeabilization probability. However, our fluorescence analysis of smooth pillars did not show such an increase, displaying permeabilization in very few cases that we have estimated to represent less than 10% (30 cells over 320 in 5 different cell cultures).

The radius of curvature of sharp pillars ($R_{sharp} \sim 20 \pm 5$ nm) falls in the range of values where, according to the simulation results reported in Figure 4c, we expect a significant decrease in tensile strength for rupture. This can explain the increase of permeabilization, as shown in Figure 5c by fluorescence images.

For sake of clarity we remark that the novelty here is represented by the role of the sharp edge and not by a fine tuning of the aspect ratio. Similar findings were also recently observed by another group that studied plasmid transfection driven by cells interfacing with 3D nanostructures [30].

We consider these results of great importance in the field for two reasons: i) Cell membrane permeabilization is reported at pillar sizes where permeabilization is not theoretically expected [26]. ii) The comparison between sharp and smooth pillars highlights the importance of membrane curvature and its role as key player in the permeabilization processes.

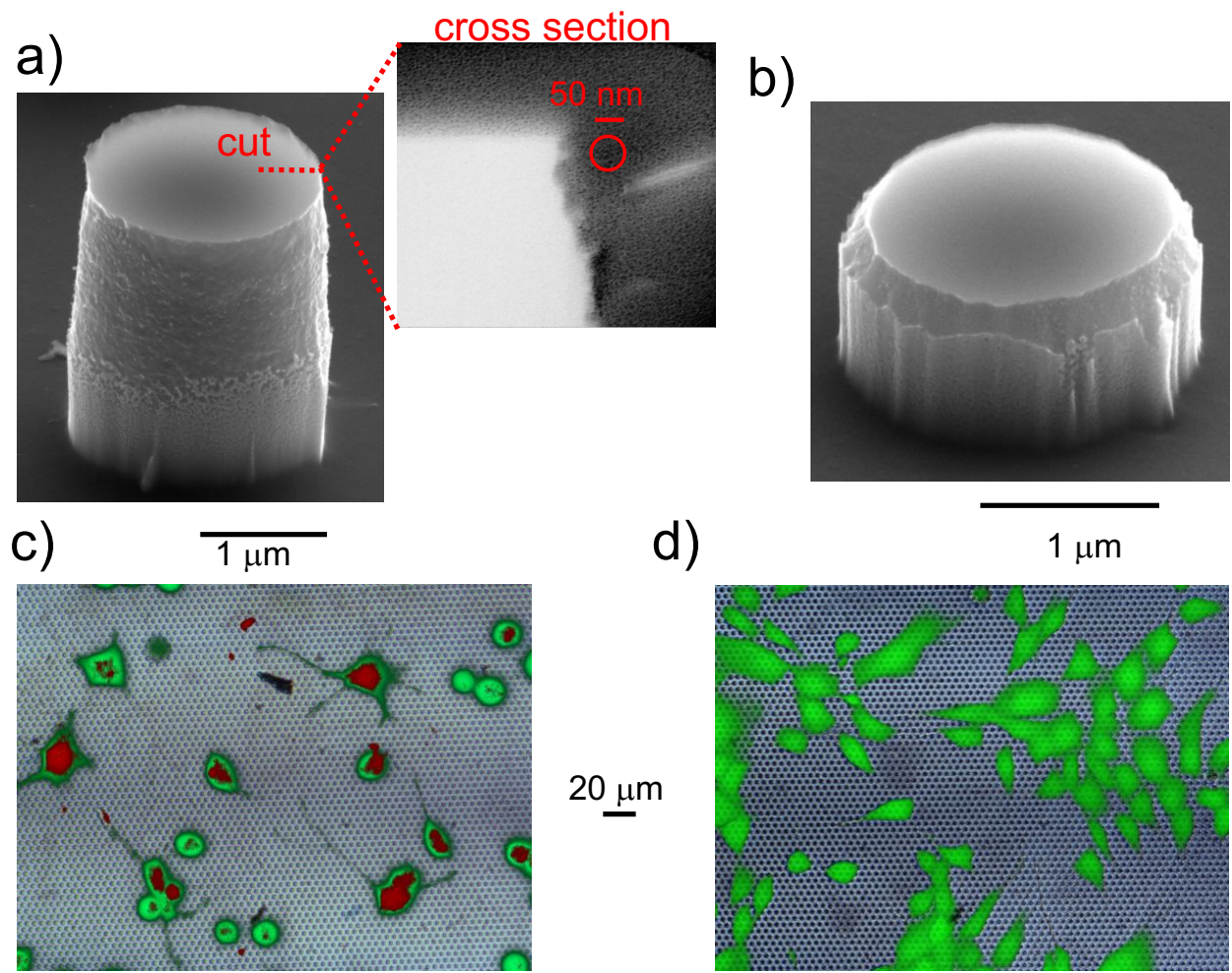

**Figure 5 a)** SEM image of a "sharp" pillar with diameter of 2 μm, height of 2.5 μm. The inset highlights a cross section of the sharp edge, with an estimated radius of curvature of $R_{sharp} \sim 20 \pm 5$ nm (compare the approximating circle with the scale bar). **b)** SEM image of a "smooth" pillar with diameter of 2 μm, height of 1 μm. The estimated radius of curvature is $R_{smooth} \sim 250 \pm 20$ nm. **c)** Fluorescence image of cells cultured on sharp pillars fabricated with spacing of 5 μm (green spots) and treated with both permeable calcein AM (green) and impermeable dye propidium iodide (red) administered in solution. Most of the cells present green and red staining, with a permeability likelihood very close to 70%. **d)** Corresponding fluorescence image of cells cultured on smooth pillars and treated with the same dyes. In this case, only a limited number of cells contain propidium iodide and are hence stained in red, meaning that the cell membrane is successfully permeabilized in a few cases (see S2 for more details). A specific red staining is probably due to fraction of death cells and DNA dispersed in the cell culture.

## CONCLUSIONS

In summary, we studied, through a 2D coarse-grained molecular dynamics model, the mechanical behavior of a cell membrane interacting with micro- and nanopillars. To identify the stress-strain relations and the rupture conditions, we performed simulations of a lipid bilayer subjected to different loading conditions: longitudinal traction, pure bending and combined traction and bending. We have also proposed a simple mechanical model for estimating the main intensive elastic parameters, such as the Young's modulus and the ultimate tensile stress, in a regime of large stress and deformation. Notably, we did not introduce any restrictive hypothesis on the constitutive relation for the membrane's elasticity: actually, the stress-strain relation could be non-linear. Moreover, we extracted the ultimate tensile loads from the simulations without postulating any value a priori. We found that the bending of a membrane is characterized by an elastic regime at low bending angles, followed by a plastic one at higher angles, namely, spatial configurations that are irreversible within the time window of the simulations.

Importantly, the simulation shows that bending of the membrane (e.g., at the nanopillar's edge) dramatically lowers the rupture force.

To support these theoretical findings, we investigated experimentally the permeabilization of cells cultured on micropillars. Surprisingly, we found that spontaneous permeabilization events may occur on pillars of diameters in the micrometer range, which is much larger than previously observed. A comparison between cells cultured on pillars with a comparable aspect ratio but different sharpness at the edges suggests that the local curvature of the membrane could be responsible for the strong increase in permeabilization, with a probability that increases from 10% in the smooth case to the 70% in the sharp case.

Therefore, the simulation results suggest a direct interpretation of the experimental data. In principle, even step edges can cause local membrane disruption provided that they are sharp enough.

The results obtained from our simulations could also be applied to the understanding of the internalization of substrate-free nanowires [42, 43], since cellular traction forces exist in that situation as well. In fact, the internalization process often starts at the sharp tip of nanowires, where the curvature is higher.

In conclusion, local geometric features at the cell-substrate interface can dramatically affect cell permeabilization to a greater extent than the effect of the pillar aspect ratio and spacing. These results may provide important information for the field of biointerfaces and tissue engineering, offering valuable insight into designing devices for gene transfection, intracellular delivery and electrical recording.

**METHODS**

**Simulation details.** We performed Molecular Dynamics simulations of a lipid bilayer in water by using Langevin dynamics that allows to calculate not only the forces due to bending, but also to the thermal fluctuations. Periodic boundary conditions were imposed in all directions and the temperature was kept at 301.7 K by means of a Langevin thermostat applied to all particles.

We assume the existence of three types of particles: "water-like" particles (w), hydrophilic "head" particles (h), and hydrophobic "tail" particles (t). A water molecule is approximated as a single "w" particle, while a lipid molecule is composed of one "h" particle and five "t" particles joined together by harmonic interactions. The three types of particles interact by means of Lennard-Jones potentials with the depth of the potential well $\epsilon_{ij} = 6.94 \cdot 10^{-21} J$, except for $\epsilon_{tt} = 11.8 \cdot 10^{-21} J$ and $\lambda_{ij} = \lambda = 0.4456 \, nm$, with $\lambda$ distance at which the inter-particle potential is 0 (see S3 for more details). All interactions are truncated at $R_{ij}^c = 2.5\lambda$, except for w-t and h-t interactions where $R_{ij}^c = 2^{1/6}\lambda = 0.5 nm$, which makes the latter interactions repulsive.

When randomly dispersed in water, the lipids undergo a self-assembly process because of the assigned interactions, eventually forming (after t=10 ns) a lipid bilayer with thickness of $l_y \approx 4.5 nm$ and average distance between the hydrophilic heads $d_{eq} \approx 0.83\ nm$. After the bilayer formation, we replicate the structure to create a system size of about 100 nm corresponding to a straight and stable membrane.

The equilibrium curvature of the membrane can be predicted estimating the two-dimensional "Shape factor" [37], $S_f = A_c/(a_h \cdot l_t)$, where $A_c$ is the area of hydrocarbon chains, $a_h$ is the head thickness, and $l_t$ is the tail length. In our case $S_f \approx 1$ and we have indeed a bilayer (see S3 for more details).

All simulations have been executed using the LAMMPS package for large scale molecular dynamics simulations. Configurations were stored every 3 ps for further structural and dynamic analyses.

**Fabrication of pillars.** For the fabrication of the silicon pillars arrays, we used first photolithography to pattern a chrome hard mask on a silicon wafer substrate and then dry etching to define the pillars in the silicon bulk using a customized multi-step process to control the shape of the pillars' top edges (see S1 for technical details).
The fabrication of the silicon pillar arrays has been done using a photolithographic chrome hard mask on a silicon wafer substrate. Silicon pillars have been fabricated by dry etching in an ICP-RIE reactor (Sentech SI 500) using a customized multi-step silicon etching process based on SF6 and C4F8 gases. ICP source has been set at 1200 W, RF power at 20 Wand reactor pressure at 1.33 Pa.
After dry etching, any residual of chrome has been removed using Chrome Etch 18.
More details on the fabrication process can be found in the Supplementary Information (S1).

**Cell culture.** The cells that appear in Figures 1 and 5 are NIH-3T3 plated with a density of $1,5 \times 10^4$ cell/cm$^2$ in DMEM complemented with 1% penicillin streptomycin antibiotics and 10% inactivated fetal bovine serum (FBS); cells are incubated at controlled temperature, humidity and $CO_2$ concentration for 48h. A viability test has been performed and both calcein AM and propidium iodide (2μM) have been administer to the cultures (see S2 for more details). After 20 minutes of incubation, the samples have been washed three times with PBS and then analyzed through a fluorescence microscope.

**Cell imaging.** Figures 1a and 1b are SEM images of a NIH-3T3 cell cross section. The cells have been fixed and stained using a recently implemented RO – T – O procedure [35] and samples have been infiltrated increasing concentration of resin in ethanol. The excess of resin has been removed prior to polymerization in order to allow the detection of the cells on the 3D nanostructures using SEM [35].

FIB/SEM cross section images have been made with a FEI Helios Nanolab 650 dual beam microscope. A 1μm layer of Pt has been deposited on the selected region by a Gas Injection System (GIS) integrated to the microscope before the cutting. An initial cut was made with a high current (9.3nA) of the ionic beam, and then a cleaning cross section was performed using a 0.79 nA ion current with a 30.0kV accelerating voltage. The imaging was made using the immersion mode of the microscope and backscattered electrons have been detected with TLD detector.

**Corresponding Author**

*Francesco De Angelis, E-mail: francesco.deangelis@iit.it, Via Morego 30, 16163 Genoa, Italy.

**Author Contributions**

R. C. conceived and performed the MD simulations, the data analysis and wrote the manuscript. V. C. carried out the experiments and associated data analysis. C. A. G. performed the mechanical and mathematical analysis of simulation data. A. B. and F. M. fabricated the nanopillars to be used for cell culture. F. S. assisted in the experimental analysis and imaging. F. D. A. conceived the experiment, supervised the work, analyzed data. All authors contributed to the preparation of the manuscript and have given approval to the final version.

**Notes**

The authors declare no competing financial interest.

ACKNOWLEDGMENTS

This work was supported by the European Research Council under the European Union's Seventh Framework Programme (FP/2007-2013) / ERC Grant Agreement no. (616213), CoG: Neuro-Plasmonics.

ASSOCIATED CONTENT

**Supporting Information**

S1: Fabrication method

S2: Cell culture and imaging

S3. MD simulation of Membrane self-assembly

S4. Strain-stress curve

S5. Bending stiffness

S6. Maximum stress on the membrane's section

S7. Stress-strain relation for bent membrane

# References [34] [34]

[1] M. P. Stewart, A. Sharei, X. Ding, G. Sahay, R. Langer and K. F. Jensen, "Break and enter: in vitro and ex vivo strategies for intracellular delivery," *Nature*, vol. 538, p. 183–192, 2016.

[2] P. Washbourne and A. K. McAllister, "Techniques for gene transfer into neurons," *Curr. Opin. Neurobiol.*, vol. 12, p. 566–573, 2002.

[3] G. He, H.-J. Chen, D. Liu, Y. Feng, C. Yang, T. Hang, J. Wu, Y. Cao and X. Xie, "Fabrication of Various Structures of Nanostraw Arrays and Their Applications in Gene Delivery," *Adv. Mater. Interfaces*, p. 1701535, 2018.

[4] Y. Ma, R. J. M. Nolte and J. J. L. M. Cornelissen, "Virus-based nanocarriers for drug delivery," *Adv. Drug Deliv. Rev.*, vol. 64, p. 811–825, 2012.

[5] C. Xie, Z. Lin, L. Hanson, Y. Cui and B. Cui, "Intracellular recording of action potentials by nanopillar electroporation," *Nat. Nanotechnol.*, vol. 7, p. 185–190, 2012.

[6] J. Teissié, N. Eynard, B. Gabriel and M. Rols, "Electropermeabilization of cell membranes," *Adv. Drug Deliv. Rev.*, vol. 35, p. 3–19, 1999.

[7] S. Yoon, M. G. Kim, C. T. Chiu, J. Y. Hwang, H. H. Kim, Y. Wang and K. K. Shung, "Direct and sustained intracellular delivery of exogenous molecules using acoustic-transfection with high frequency ultrasound," *Sci. Rep.*, vol. 6, p. 20477, 2016.

[8] Q. Fan, W. Hu and A. T. Ohta, "Efficient single-cell poration by microsecond laser pulses," *Lab Chip,* vol. 15, p. 581–588, 2015.

[9] R. Xiong, S. K. Samal, J. Demeester, A. G. Skirtach, S. C. De Smedt and K. Braeckmans, "Laser-assisted photoporation: fundamentals, technological advances and applications," *Adv. Phys. X,* vol. 1, no. 4, p. 596–620, 2016.

[10] N. Saklayen, M. Huber, M. Madrid, V. Nuzzo, D. I. Vulis, W. Shen, J. Nelson, A. A. McClelland, A. Heisterkamp and E. Mazur, "Intracellular Delivery Using Nanosecond-Laser Excitation of Large-Area Plasmonic Substrate," *ACS Nano,* vol. 11, p. 3671–3680, 2017.

[11] A. R. Durney, L. C. Frenette, E. C. Hodvedt, T. D. Krauss and H. Mukaibo, " Fabrication of Tapered Microtube Arrays and Their Application as a Microalgal Injection Platform," *ACS Appl. Mater. Interfaces,* vol. 8, p. 34198–34208, 2016.

[12] C. Chiappini, J. O. Martinez, E. De Rosa, C. S. Almeida, E. Tasciotti and M. M. Stevens, "Biodegradable Nanoneedles for Localized Delivery of Nanoparticles in Vivo: Exploring the Biointerface," *ACS nano,* vol. 9, p. 5500, 2015.

[13] C. Chiappini, E. De Rosa, J. O. Martinez, X. Liu, J. Steele, M. M. Stevens and E. Tasciotti, "Biodegradable silicon nanoneedles delivering nucleic acids intracellularly induce localized in vivo neovascularization," *Nature Mat.,* vol. 14, p. 532, 2015.

[14] A. Hai, J. Shappir and M. E. Spira, "In-cell recordings by extracellular microelectrodes," *Nature Methods,* vol. 7, p. 200, 2010.

[15] A. Hai and M. E. Spira, "On-chip electroporation, membrane repair dynamics and transient in-cell recordings by arrays of gold mushroom-shaped microelectrodes," *Lab Chip,* vol. 12, p. 2865, 2012.

[16] V. Caprettini, A. Cerea, G. Melle, L. Lovato, R. Capozza, J.-A. Huang, F. Tantussi, M. Dipalo and F. De Angelis, "Soft electroporation for delivering molecules into tightly adherent mammalian cells through 3D hollow nanoelectrodes," *Sci. Rep.,* vol. 7, p. 8524, 2017.

[17] A. K. Shalek, J. T. Robinson, E. S. Karp, J. S. Lee, D.-R. Ahn, M.-H. Yoon, A. Sutton, M. Jorgolli, R. S. Gertner, T. S. Gujral, G. MacBeath, E. G. Yang and H. Park, "Vertical silicon nanowires as a universal platform for delivering biomolecules into living cells," *PNAS,* vol. 107, p. 1870, 2010.

[18] X. Xie, A. Aalipour, S. Gupta and N. Melosh, "Determining the Time Window for Dynamic Nanowire Cell Penetration Processes," *ACS Nano,* vol. 9, p. 11667–11677, 2015.

[19] Z. C. Lin, C. Xie, Y. Osakada, Y. Cui and B. Cui, Iridium oxide nanotube electrodes for sensitive and prolonged intracellular measurement of action potentials, vol. 5, Nat. Comm., 2014, p. 3206.

[20] M. M. Stevens and J. H. George, "Exploring and Engineering the Cell Surface Interface," *Science,* vol. 310, p. 1135, 2005.

[21] J. Huang, S. Grater, F. Corbellini, S. Rinck, E. Bock, R. Kemkemer, H. Kessler, J. Ding and J. Spatz, "Impact of Order and Disorder in RGD Nanopatterns on Cell Adhesion," *Nano Lett.,* vol. 9, p. 1111, 2009.

[22] H. Persson, Z. Li, J. O. Tegenfeldt, S. Oredsson and C. N. Prinz, "From immobilized cells to motile cells on a bed-of-nails: effects of vertical nanowire array density on cell behaviour," *Sci. Rep.,* vol. 5, p. 18535, 2015.

[23] M. Toma, A. Belu, D. Mayer and A. Offenhäusser, "Flexible Gold Nanocone Array Surfaces as a Tool for Regulating Neuronal Behavior," *small*, p. 1700629, 2017.

[24] A. Sharei, J. Zoldan, A. Adamo, W. Y. Sim, N. Cho, E. Jackson, S. Mao, S. Schneider, M.-J. Han, A. Lytton-Jean, P. A. Basto, S. Jhunjhunwal, J. Lee, D. A. Heller, J. W. Kang, G. C. Hartoularos, K.-S. Kim, D. G. Anderson, R. Langer and K. F. Jensen, "A vector-free microfluidic platform for intracellular delivery," *PNAS ,* vol. 110, p. 2082, 2013.

[25] W. Zhao, L. Hanson, H.-Y. Lou, M. Akamatsu, P. D. Chowdary, F. Santoro, J. R. Marks, A. Grassart, D. G. Drubin, Y. Cui and B. Cui, "Nanoscale manipulation of membrane curvature for probing endocytosis in live cells," *Nat. Nanotech.,* vol. 12, p. 750, 2017.

[26] X. Xie, A. M. Xu, M. R. Angle, N. Tayebi, Verma, P. and N. A. Melosh, "Mechanical Model of Vertical Nanowire Cell Penetration," *Nano Lett.,* vol. 13, p. 6002–6008, 2013.

[27] W. Kim, J. K. Ng, M. E. Kunitake, B. R. Conklin and P. Yang, "Interfacing Silicon Nanowires with Mammalian Cells," *J. Am. Chem. Soc.,* vol. 129, p. 7228–7229, 2007.

[28] R. Yan, J.-H. Park, Y. Choi, C.-J. Heo, S.-M. Yang, L. P. Lee and A. P. Yang, "Nanowire-based single-cell endoscopy," *Nat. Nanotechnol.,* vol. 7, p. 191–196, 2012.

[29] R. Singhal, Z. Orynbayeva, R. V. K. Sundaram, J. J. Niu, S. Bhattacharyya, E. A. Vitol, M. G. Schrlau, E. S. Papazoglou, G. Friedman and Y. Gogotsi, "Multifunctional carbon-nanotube cellular endoscopes," *Nat. Nanotechnol.,* vol. 6, p. 57–64, 2011.

[30] J. Harding F., S. Surdo, B. Delalat, C. Cozzi, R. Elnathan, S. Gronthos, H. Voelcker N. and G. Barillaro, "Ordered Silicon Pillar Arrays Prepared by Electrochemical Micromachining: Substrates for High-Efficiency Cell Transfection," *ACS Appl. Mater. Interfaces,* vol. 8, p. 29197–29202, 2016.

[31] C. H. Huang, P. Hsiao, F. Tseng, S. Fan, C. Fu and R. Pan, "Pore-Spanning Lipid Membrane under Indentation by a Probe Tip: a Molecular Dynamics Simulation Study," *Langmuir,* vol. 27, pp. 11930-42, 2011.

[32] S. Kawamoto, T. Nakamura, S. O. Nielsen and W. Shinoda, "A guiding potential method for evaluating the bending rigidity of tensionless lipid membranes from molecular simulation," *J. Chem. Phys.*, vol. 139, p. 034108, 2013.

[33] M. Hu, P. Diggins and M. Deserno, "Determining the bending modulus of a lipid membrane by simulating buckling," *J. Chem. Phys.*, vol. 138, p. 214110, 2013.

[34] H. Noguchi, "Anisotropic surface tension of buckled fluid membranes," *Phys. Rev E*, vol. 83, p. 061919, 2011.

[35] F. Santoro, W. Zhao, L.-M. Joubert, L. Duan, J. Schnitker, Y. van de Burgt, H.-Y. Lou, B. Liu, A. Salleo, L. Cui, Y. Cui and B. Cui, "Revealing the Cell–Material Interface with Nanometer Resolution by Focused Ion Beam/ Scanning Electron Microscopy," *ACS Nano*, vol. 11, p. 8320, 2017.

[36] M. Venturoli, M. M. Sperotto, M. Kranenburg and B. Smit, "Mesoscopic models of biological membranes," *Physics Reports*, vol. 437, p. 1 – 54, 2006.

[37] J. N. Israelachvili, Intermolecular and Surface Forces, London: Academic Press, 2011.

[38] R. Goetz, G. Gompper and R. Lipowsky, "Mobility and Elasticity of Self-Assembled Membranes," *Phys. Rev. Lett.*, vol. 82, p. 221, 1999.

[39] E. Lindahl and O. Edholm, "Mesoscopic Undulations and Thickness Fluctuations in Lipid Bilayers from Molecular Dynamics Simulations," *Biophys. J.*, vol. 79, p. 426, 2000.

[40] G. Brannigan, L. Lin and F. Brown, "Implicit solvent simulation models for biomembranes," *Eur. Biophys. J.*, vol. 35, p. 104, 2006.

[41] S. J. Marrink and A. E. Mark, "Effect of Undulations on Surface Tension in Simulated Bilayers," *J. Phys. Chem. B*, vol. 105, p. 6122, 2001.

[42] J. F. Zimmerman, R. Parameswaran, G. Murray, Y. Wang, M. Burke and B. zimm, "Cellular uptake and dynamics of unlabeled freestanding silicon nanowires," *Sci. Adv.*, vol. 2, p. 1601039, 2016.

[43] J.-H. Lee, A. Zhang, S. S. You and C. M. Lieber, "Spontaneous Internalization of Cell Penetrating Peptide-Modified Nanowires into Primary Neurons," *Nano Lett.*, vol. 16, p. 1509, 2016.

# GRAPHICAL ABSTRACT

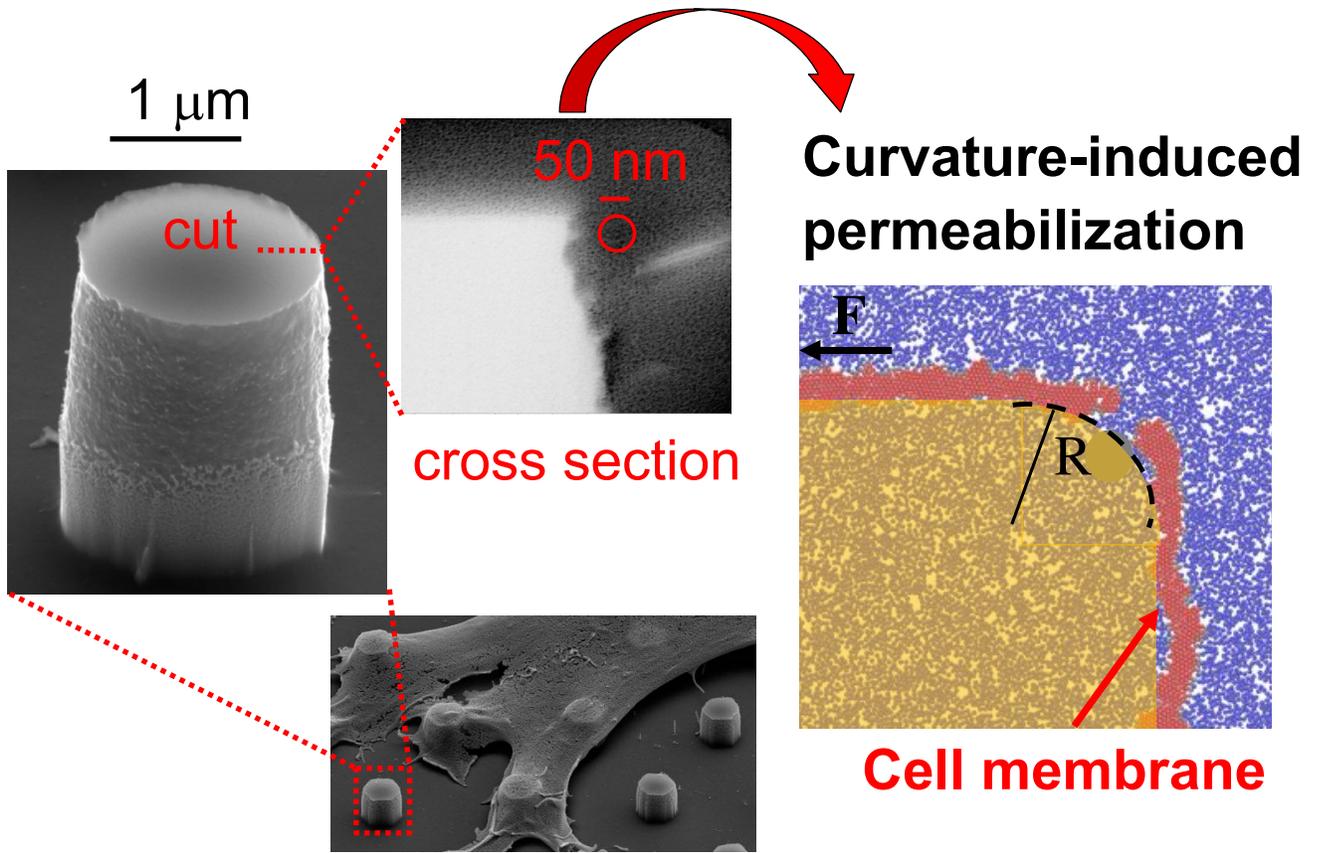